\documentclass[letterpaper]{article}
\usepackage{iccc}

\usepackage{times}
\usepackage{amsmath}
\usepackage{amssymb}
\usepackage{helvet}
\usepackage{courier}
\usepackage{multirow}
\usepackage{xcolor}
\usepackage{graphicx}
\usepackage{bm}
\title{LyricJam: A system for generating lyrics for live instrumental music}
\author{Olga Vechtomova, Gaurav Sahu, Dhruv Kumar \\ University of Waterloo \\ \texttt{\{ovechtom, gsahu, d35kumar\}@uwaterloo.ca}}

\begin{document} 
\maketitle
\begin{abstract}
\begin{quote}
We describe a real-time system that receives a live audio stream from a jam session and generates lyric lines that are congruent with the live music being played.  Two novel approaches are proposed to align the learned latent spaces of audio and text representations that allow the system to generate novel lyric lines matching live instrumental music.  One approach is based on adversarial alignment of latent representations of audio and lyrics, while the other approach learns to transfer the topology from the music latent space to the lyric latent space. A user study with music artists using the system showed that the system was useful not only in lyric composition, but also encouraged the artists to improvise and find new musical expressions. Another user study demonstrated that users preferred the lines generated using the proposed methods to the lines generated by a baseline model.
\end{quote}
\end{abstract}

\section{Introduction}

Music artists have different approaches to writing song lyrics. Some write lyrics first, and then compose music, others start with the music and let the mood and the emotions that emerge from the music guide their choice of lyric themes, words and phrases. The latter type of songwriting is often experimental in nature, and is commonly found in genres such as rock, jazz and electronic music, whereby one or more artists jam and in the process converge on the desired musical expressions and sounds. In this work, we describe a system we designed to support this type of songwriting process, where artists play musical instruments, while the system listens to the audio stream and generates lyric lines that are congruent with the music being played in real time. The artists see the lines as they are generated in real time,  potentially entering a feedback loop, where the lines suggest emotions and themes that artists can use to guide their musical expressions and instrumentation as they jam. The generated lines are not intended to be the complete song lyrics, instead acting as snippets of ideas and expressions inspiring creativity. After the jam session is over, the lines are saved, and the artist can use them as inspiration to write the lyrics for their song.

The goal of our system is to generate lyric lines that are congruent with emotions and moods evoked by live instrumental music. Past research in musicology has found a correlation between some aspects of music and emotions. In one large-scale study, researchers found evidence that certain harmonies have strong associations with specific emotions, for example, the diminished seventh is associated with the feeling of despair \cite{willimek2014minor}. In another study, researchers found correlation between major chords and lyric words with positive valence \cite{kolchinsky2017minor}. In addition to chords and harmonies, various sound textures and effects also contribute to the emotional intent of the music. Therefore, in this work we use raw audio input that captures all aspects of music. Our model learns associations between raw audio clips and their corresponding lyrical content from the training dataset, and at inference time, relies on these learned associations to generate new lines based on the live audio stream.

Our approach is based on training a variational autoencoder for learning the representations of mel-spectrograms of audio clips (spec-VAE), and a conditional variational autoencoder for learning the representations of lyric lines (text-CVAE). The advantage of using variational autoencoders as generative models is their ability to learn a continuous latent space that can then be sampled to generate novel lines, which is an important requirement for creative applications.

At inference time, the model must be able to generate new lyric lines given an audio clip being recorded from live jam session. In order to do that we need a way to align the latent representations learned by the spec-VAE and the latent representations learned by the text-VAE. We propose two novel approaches to achieve this alignment. 

\begin{figure*}[!t]
	\centering \small
	\includegraphics[width=.85\linewidth]{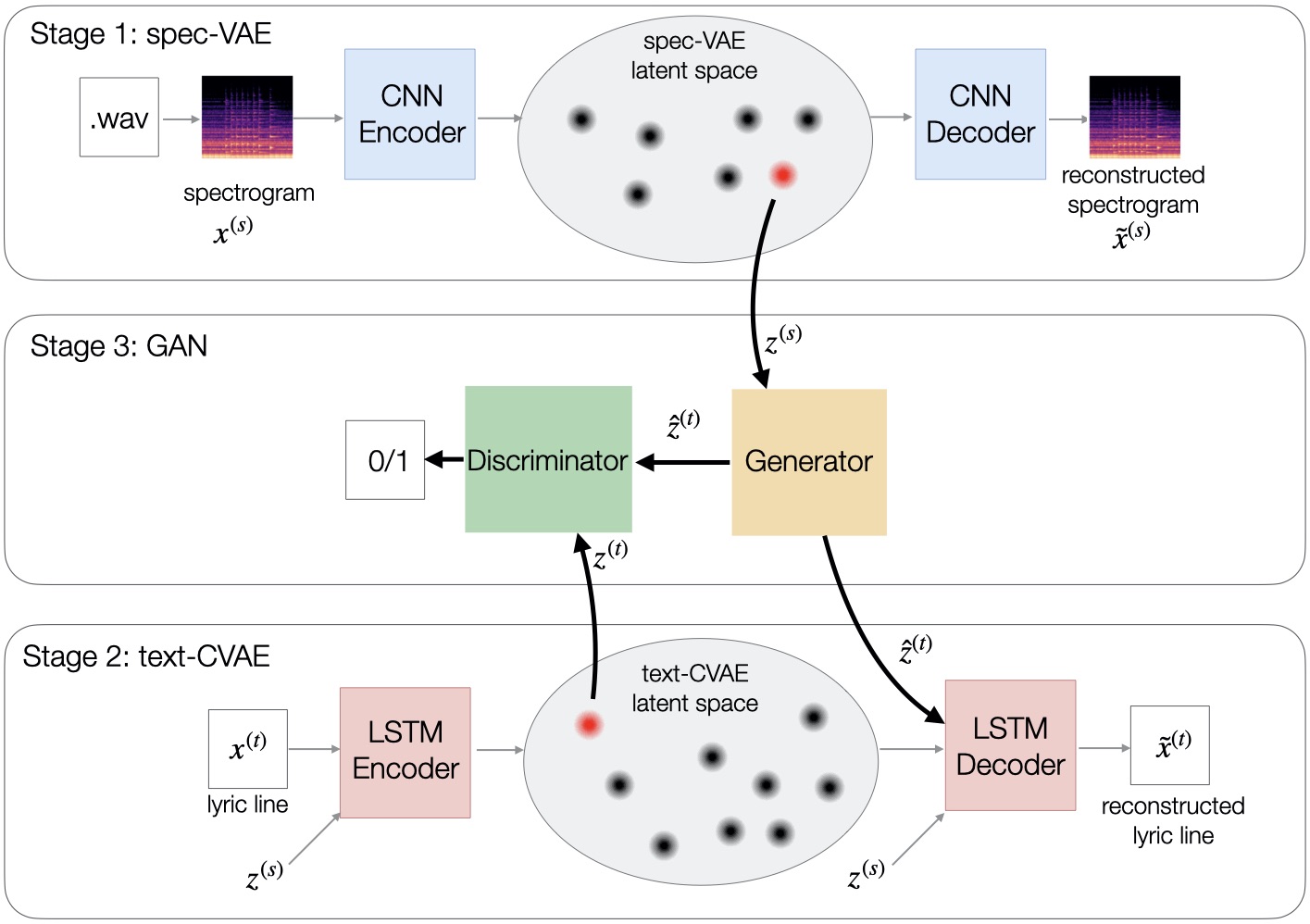}
	\caption{GAN-based alignment of music and lyrics representations (Approach 1).}

	\label{fig:GAN}
\end{figure*}

The first approach (Figure~\ref{fig:GAN}) is based on training a separate Generative Adversarial Network (GAN) model that takes the spectrogram embedding from spec-VAE, and learns to predict the lyric line embedding in the text-CVAE. The GAN-predicted embedding is then sent to the text-CVAE decoder to generate text.

\begin{figure*}[!t]
	\centering \small
	\includegraphics[width=.85\linewidth]{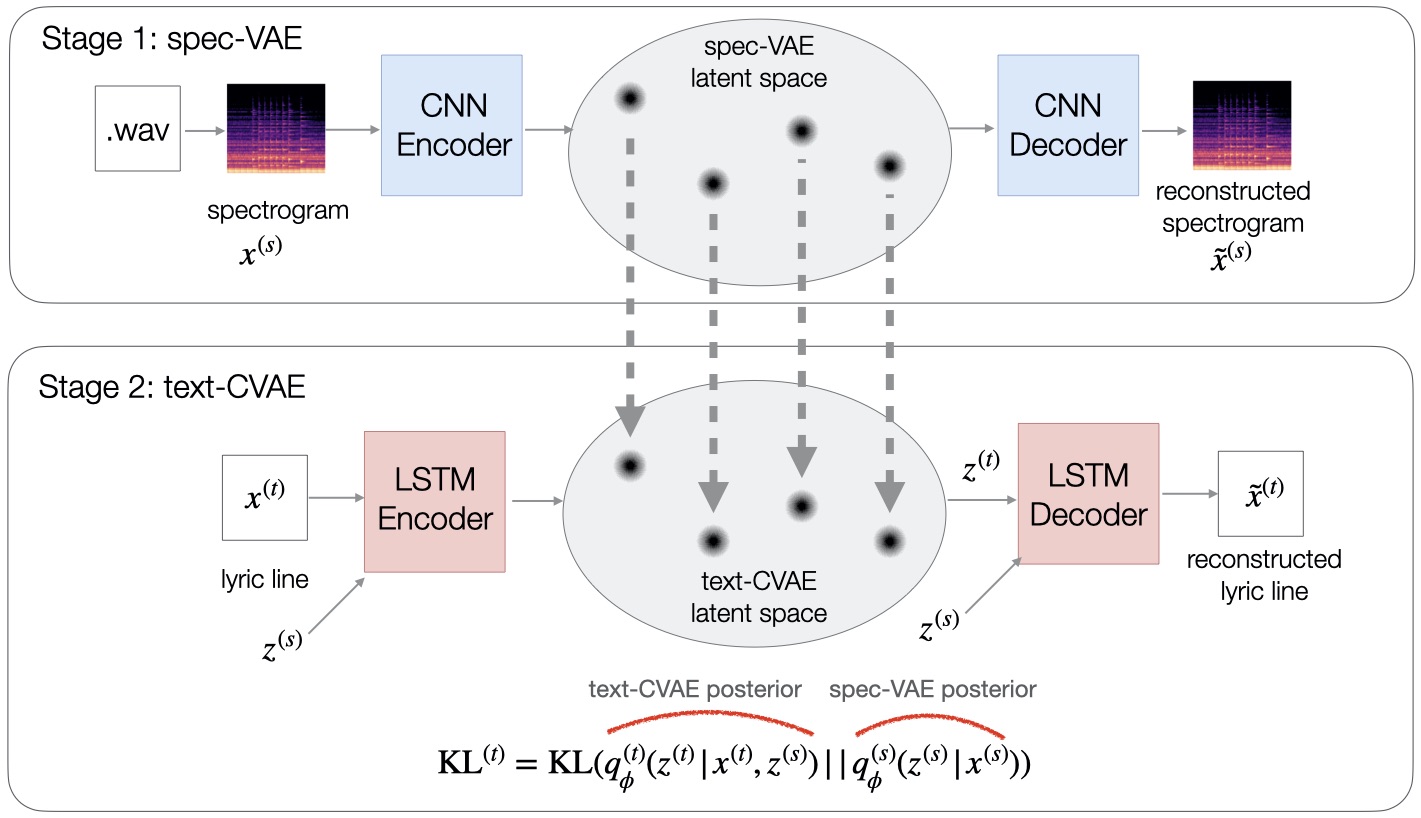}
	\caption{Latent space topology transfer from spec-VAE to text-CVAE (Approach 2).}
	\label{fig:spec-CVAE}
\end{figure*}

The second approach (Figure~\ref{fig:spec-CVAE}) learns to transfer the latent space topology of the spec-VAE to the text-CVAE latent space. To achieve this we use the learned posteriors from the spec-VAE as priors in text-CVAE during its training. The text-CVAE learns to encode lyric lines corresponding to a given audio clip in the region of the latent space corresponding to that clip. Also, since similar sounding audio clips are encoded in neighboring regions, the text-CVAE correspondingly learns to encode lines for these clips in neighboring regions. For example, ambient music clips would be encoded in proximal regions of spec-VAE, and so would be the lines corresponding to these clips. The intuition is that lines corresponding to similar sounding audio clips (e.g. ambient) would have similar emotional intent, as opposed to, for example, aggressive sounding music. At inference time, when an artist plays an ambient music piece, the system would feed its spectrogram to the spec-VAE encoder to get the parameters of its posterior distribution. Since the spec-VAE posterior distributions are also prior distributions in the text-CVAE, the system samples latent codes from the corresponding prior of the text-CVAE, generating new lines reflecting ambient music.

To summarize the contributions of this paper are as follows:
\begin{enumerate}
\item To our knowledge this is the first real-time system that receives a live audio stream from a jam session and generates lyric lines that are congruent with the live music being played.
\item We propose two novel approaches to align the latent spaces of audio and text representations that allow the system to generate novel lyric lines matching the live music audio clip. 
\item We discuss our findings based on observations and interviews of musicians using the system.
\end{enumerate}

In the next section, we describe the GAN-based approach of aligning the spec-VAE and text-VAE latent embeddings. Next, we will describe our second approach of aligning the topologies of the two VAEs. We will then follow up with the implementation details, experiments and user studies.

\section{Approach 1: GAN-based alignment of music and lyrics representations}
The model consists of three neural network models that are trained consecutively in three stages (Figure~\ref{fig:GAN}).

\subsection{Training stage 1: spec-VAE}
In this stage, we train the spectrogram variational autoencoder (VAE) model to learn the latent representations of audio clips.

First we convert the raw waveform audio files into MEL-spectrogram images using the same method as used in \citeauthor{vechtomovaLyrics2020}~\shortcite{vechtomovaLyrics2020}. These spectrograms are then used as input for the spec-VAE.

The variational autoencoder \cite{Kingma2014} is a stochastic neural generative model that consists of an encoder-decoder architecture. The encoder transforms the input image $\bm x$ into the approximate posterior distribution $q_\phi(z|x)$ learned by optimizing parameters $\phi$ of the encoder. The decoder reconstructs $\bm x$ from the latent variable $\bm z$, sampled from $q_\phi(z|x)$. In our implementation, we use convolutional layers as the encoder and a deconvolutional layers as the decoder. Standard normal distribution was used as the prior distribution $p(z)$. The VAE is trained on the loss function that combines the reconstruction loss (Mean Squared Error) and KL divergence loss that regularizes the latent space by pulling the posterior distribution to be close to the prior distribution.

\subsection{Training stage 2: text-CVAE}
Unlike the vanilla VAE used for encoding spectrograms, we use conditional VAE (CVAE) for encoding lyrics. 

The CVAE learns a posterior distribution that is conditioned not only on the input data, but also on a class c: $q_\phi(z|x,c)$. Here, we define the class as the spectrogram corresponding to a given line. Every conditional posterior distribution is pulled towards the same prior, here the standard normal distribution. 

During training, every input data point consists of a lyric line and its corresponding spectrogram. We first pass the spectrogram through the spec-VAE encoder to get the parameters of the posterior distribution (a vector of means and a vector of standard deviations). We then sample from this posterior to get a vector $z^{(s)}$ that is then concatenated with the input of the encoder and the decoder during training. The reason why we used sampling as opposed to the mean $z^{(s)}$ vector is to induce the text-VAE model to learn conditioning on continuous data, as opposed to discrete classes. This prepares it to better handle conditioning on unseen new spectrograms at inference. 

Both the encoder and the decoder in the text-CVAE are Long Short Memory Networks (LSTMs). The sampled $z^{(s)}$ is concatenated with the word embedding input to every step of the encoder and the decoder.

\subsubsection{Training stage 3: GAN}
In this phase, we train a generative adversarial network (GAN), which learns to align audio and text latent codes.
The GAN architecture has a generator $\mathit{G}$ and a discriminator $\mathit{D}$.
For training the GAN on a given spectrogram-text pair $\{x^{(s)}, x^{(t)}\}$\footnote{Superscripts $^{(s)}$ and $^{(t)}$ in our notation refer to spectrogram and text, respectively}, we follow these steps:

\begin{enumerate}
	\item First, we pass the spectrogram $x^{(s)}$ through spec-VAE to obtain $z^{(s)} = \mu^{(s)} + \tau (\epsilon \cdot \sigma^{(s)})$, the latent code sampled from the posterior distribution.
	Here, $\mu^{(s)}$ and $\sigma^{(s)}$ denote the mean and standard deviation predicted by the spec-VAE, $\epsilon \sim \mathcal{N}(0, 1)$ is a random normal noise, and $\tau$ is the sampling temperature.
	Simultaneously, we obtain $z^{(t)} = \mu^{(t)} + \tau (\epsilon \cdot \sigma^{(t)})$ by passing the corresponding lyric line $x^{(t)}$ through the text-VAE.
	
	\item After obtaining $z^{(s)}$ and $z^{(t)}$, we proceed with the GAN training.
	We pass $z^{(s)}$ through the generator network, which outputs a predicted text latent code $\hat{z}^{(t)}$.
	
	\item We then pass $\hat{z} = [\hat{z}^{(t)}; z^{(s)}]$ and $z = [z^{(t)}; z^{(s)}]$ through the discriminator network, where $;$ denotes the concatenation operation.
	We treat $\hat{z}$ as the negative sample, and ${z}$ as the positive sample.
	The discriminator $\mathit{D}$ then tries to distinguish between the two types of inputs.
	This adversarial training regime, in turn, incentivizes $\mathit{G}$ to match $\hat{z}^{(t)}$ as closely as possible to $z^{(t)}$.
\end{enumerate}

The adversarial loss is formulated as follows:

\begin{equation}
\begin{split}
    \min_G \max_D V(D, G) &= \mathbb{E}_{x \sim \mathcal{D}_\text{train}}[\textrm{log} D({z}) + \textrm{log}(1 - D(\hat{z})]
\end{split}
\end{equation}

where $\mathcal{D}_\text{train}$ is the training data, and each sample $x=\{x^{(s)}, x^{(t)}\}$.
We also add an auxiliary MSE loss to the objective function as it is found to stabilize GAN training~\cite{khan-etal-2020-adversarial}.
The overall loss for the GAN is:

\begin{equation}
    J_{GAN} = \min_G \max_D V(D, G) + \lambda_{MSE} || \hat{z}^{(t)} - {z}^{(t)} ||
\end{equation}

At inference time, the encoder of the text-CVAE is no longer needed. A spectrogram is input to the spec-CVAE encoder to obtain the spectrogram latent code $z^{(s)}$, which is then fed to the generator of the GAN, which generates the lyric latent code $z^{(t)}$. The inference process is also stochastic, as $z^{(s)}$ is sampled from the posterior distribution for $s$. Sampling allows us to generate diverse lines for the same spectrogram. The GAN-predicted text latent code is then concatenated with the spectrogram latent code and input to the text-CVAE decoder, which generates a lyric line.

\section{Approach 2: Latent space topology transfer from spec-VAE to text-CVAE}
The intuition for this approach is to induce the text-CVAE to learn the same latent space topology as the spec-VAE. This would mean that data points that are close in the spec-VAE latent space are expected to be close in the text-VAE latent space. More concretely, if two audio clips are encoded in the neighboring regions of the spec-VAE latent space, their corresponding lyric lines should be encoded in the neighboring regions in the text-CVAE latent space. 

The training is a two-stage process (Figure~\ref{fig:spec-CVAE}), where the first stage, spec-VAE, is the same as in the GAN-based approach. For the second stage, we train a different formulation of text-CVAE. Instead of using one prior (standard normal) to regularize every posterior distribution, we use the posterior of the spec-VAE as the prior for any given data point. More formally, let the spectrogram be $x^{(s)}$ and its corresponding lyric line be $x^{(t)}$. The posterior distribution for the spectrogram in the spec-VAE is $q^{(s)}_\phi(z^{(s)}|x^{(s)})$, and the posterior distribution for the lyric line in the text-CVAE is $q^{(t)}_\phi(z^{(t)}|x^{(t)}, z^{(s)})$. The KL term of the text-CVAE loss is computed between the posterior for the lyric line and the prior which is set to be the posterior of its corresponding spectrogram in spec-VAE:

\begin{equation}
\begin{split}
\operatorname{KL}^{(t)} = \operatorname{KL}(q^{(t)}_\phi(z^{(t)}|x^{(t)}, z^{(s)})  || q^{(s)}_\phi(z^{(s)}|x^{(s)}))
\end{split}
\label{eq:rec_loss_acg}
\end{equation}

The cross-entropy reconstruction loss for text-CVAE is:

\begin{equation}
\begin{split}
    J_\text{rec}(\phi, \theta, {z}^{(s)}, x^{(t)})= -\sum_{i=1}^n \log p({x}^{(t)}|\bm {z}^{(t)},\\ 
    {z}^{(s)},
    {x_1}^{(t)}\cdots {x_{i-1}}^{(t)})
\end{split}
\label{eq:rec_loss_acg}
\end{equation}

The final text-CVAE loss is the combination of the reconstruction loss and the KL term: 

\begin{equation}
\begin{split}
    J_\text{CVAE}(\phi, \theta, {z}^{(s)}, x^{(t)})= J_\text{rec} + \lambda \operatorname{KL}^{(t)}
\end{split}
\label{eq:cvae_loss}
\end{equation}

To avoid the KL term collapse problem, we used two techniques first proposed by \citeauthor{BowmanVVDJB15}~\shortcite{BowmanVVDJB15}: KL term annealing by gradually increasing $\lambda$ and word dropout from the decoder input.

\section{Ranking of generated lines with BERT}
At inference time, the model generates a batch of 100 lines conditioned on a short audio clip sampled every 10 seconds from the user's audio source. Since the system shows one or two lines to the user for each clip, and since not all generated lines are equally fluent or interesting, we need a method to rank them so that the system shows a small number of high-quality lines to the user. For this purpose, we used a pre-trained BERT model \cite{bert2019}, which we fine-tuned on our custom dataset. The dataset consisted of 3600 high-quality and low-quality lines that were manually selected by one of the authors from a large number of lines output by a VAE trained on the same dataset as used in our experiments. The criteria used for determining quality of lines included the following: originality, creativity, poetic quality and language fluency. While BERT is trained as a binary classifier, we use logits from the final layer for ranking the lines.

\section{Live lyric generation system}
\begin{figure}[!t]
	\centering \small
	\includegraphics[width=1\linewidth]{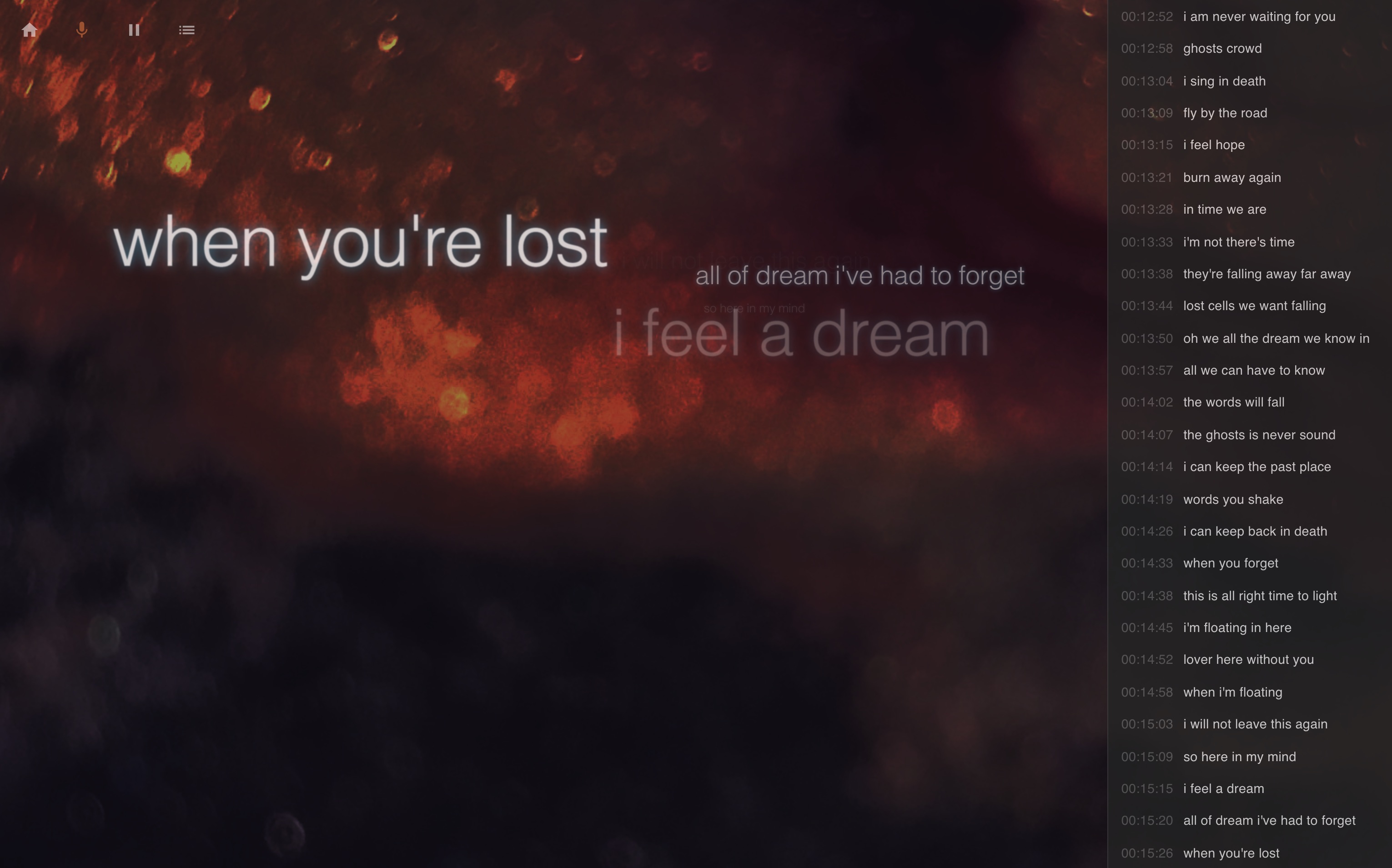}
	\caption{LyricJam screenshot.}
	\vspace{-.5cm}
	\label{fig:screenshot}
\end{figure}

We developed a React/NodeJS web application\footnote{The application can be accessed at: https://lyricjam.ai} that listens to the user's audio source, which can be either a microphone or line level input from the user's audio-to-digital converter, receiving input from the user's instruments, such as guitar or keyboard. The application samples clips from the audio stream every 10 seconds and saves them as uncompressed PCM WAV files at 44.1kHz sampling rate. On the server, WAV files are converted to MEL-spectrograms and sent to the spec-VAE to obtain the latent code, which is then used in lyric generation by the text-CVAE. The lines generated by text-CVAE are passed through BERT classifier for ranking. The top-ranked lines are then displayed to the user on their screen. The lines slowly float for a few seconds, gradually fading away as newly generated lines appear. The user can view the history of all generated lines with time stamps during the jam session in a collapsible side drawer (Figure~\ref{fig:screenshot}).

\section{Evaluation}
We trained the system on the aligned lyric-music dataset created by \citeauthor{vechtomovaLyrics2020}~\shortcite{vechtomovaLyrics2020}. The dataset consists of 18,000 WAV audio clips of original songs and their corresponding lyrics by seven music artists.

The goals of our evaluation are two-fold:
\begin{enumerate}
    \item Determine whether the two proposed methods generate lines that match the mood created by the music more accurately.
    \item Understand how the application can benefit music artists as they compose new songs by playing musical instruments.
\end{enumerate}

To answer these evaluation questions, we designed two user studies, described in detail in the subsequent sections.

\subsection{Study 1}
For this study, we developed an interactive web application, which plays instrumental songs and every 10 seconds shows three lines to the users in random order, two generated by the proposed models and one by the baseline model. The user was asked to click on the line that best matches the mood of the music currently being played. Participants for this study did not have to be songwriters or music artists, but they were required to have a general appreciation of music of any genre. 

For the baseline, we used CVAE with standard normal prior. This is a strong baseline, which generates lines conditioned on the music being played. Comparison to this baseline specifically lets us evaluate the effect of the two proposed methods: GAN-based alignment (GAN-CVAE) and topology transfer (CVAE-spec). At inference time, the lines generated by each system were ranked by BERT, and the top-ranked line from each system was selected for the experiment.

\begin{figure}
	\centering \small
	\includegraphics[width=1\linewidth]{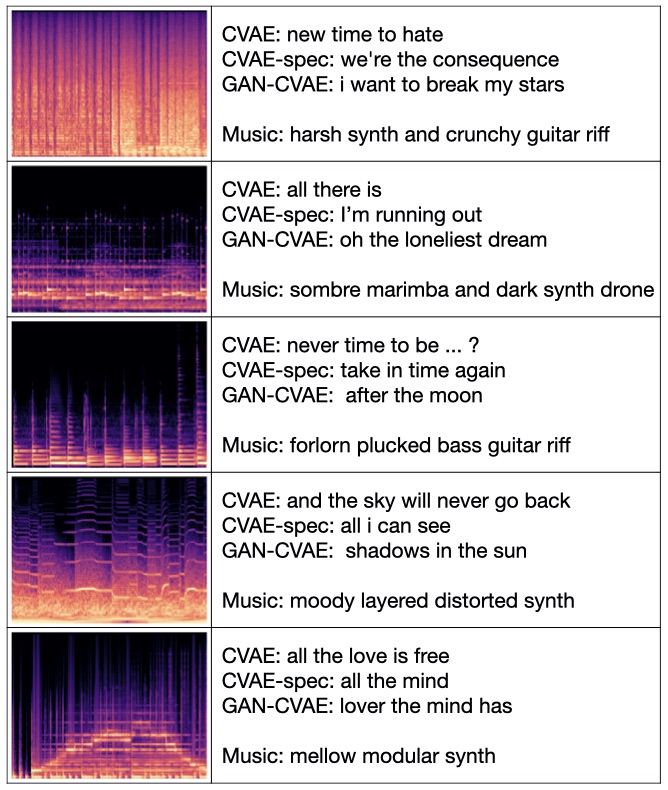}
	\caption{Examples of generated lines for different instrumental music clips.}
	\vspace{-.5cm}
	\label{fig:lyrics}
\end{figure}

We selected five instrumental songs by two artists for this study with a variety of tempos and instrumentation, and evoking different moods. The lines for each 10-second song clip were generated in advance, so that each user was shown exactly the same three lines per clip. The order of line presentation was random each time. Examples of generated lyrics can be seen in Figure~\ref{fig:lyrics}.

\begin{table*}
\small
	\centering
%	\resizebox{\linewidth}{!}{
		\begin{tabular}{ c | c | c | c }
			\hline
			\textbf{Song description} & \textbf{CVAE} & \textbf{CVAE-spec} & \textbf{GAN-CVAE}\\
			\hline
			Synthesizer, bass guitar, banjo, piano / forlorn, tense, slow & 162 & 152 & \textbf{195} \\
			\hline
			Electric guitar, synthesizer / harsh, aggressive, high-tempo & 61 & \textbf{94} & 89 \\
			\hline
			Modular synthesizer, keyboard / melancholy, low-tempo & 64 & 48 & \textbf{76} \\
			\hline
			Piano / sombre, tense, low-tempo & 43 & \textbf{91} & 66 \\
			\hline
			Modular synthesizer, keyboard / mellow, uplifting, ambient & 53	& 72 & 53\\
			\hline
			\textbf{Total} & 383 & 457 & \textbf{479}\\
			\hline
		\end{tabular}
		\vspace{1em}
	\caption{Number of lines per system and instrumental song selected by the users in Study 1}
	\label{tab:lineSelection}
\end{table*}

In total, 15 users participated in this study. As can be seen from Table \ref{tab:lineSelection}, users preferred the lines from two experimental systems over the baseline regardless of the type of song played. The differences are statistically significant (ANOVA, p$<$0.01).

\subsection{Study 2}
For this study we asked five musicians to use the system as they played musical instruments. The participants were free to choose any musical instruments they liked and play any genre of music. They were asked to observe the generated lines as they played music, and answer the following questions after the completion of the study:

\begin{enumerate}
\item What were your thoughts after using the application?
\item What did you like about it?
\item What would you change about this system?
\item Would you consider using this system again?
\item Did the lines reflect the mood that your music conveys to you?
\end{enumerate}

The instruments and genres played by the participants are as follows: 
\begin{itemize}
    \item Participant A: electric guitar / rock
    \item Participant B: keyboard / rock
    \item Participant C: guitar and electric drums / blues, rockabilly, punk, folk, rock.
    \item Participant D: piano / new age, contemporary classical, Bollywood, pop;
    \item Participant E: electric guitar / metal
\end{itemize}

All users found the experience enjoyable and would use the system again. Three users mentioned that they liked its simplicity and minimalist design. Interestingly, while our initial goal was to design a system to assist musicians in lyric writing, two unanticipated new uses emerged from this user study: (a) the system encouraged improvisation and trying new sounds and (b) the system was perceived as a helpful jam partner. This suggests that the system could be useful not only for lyric composition, but also for music playing in general. Below, we elaborate in more detail on the themes that emerged from the feedback.

\subsubsection{Improvisation and experimenting with new sounds.}

Users commented that the system encouraged them to try new sounds and experiment with new musical expressions. User A: ``[The system] inspires me to try things out to see what this sound does to the machine''. User C: ``there were a few times where the system started suggesting lyrics that got me thinking about structuring chords a bit differently and taking a song in different directions than originally intended''.

User A also commented that the system encouraged them to be more experimental in their music: ``You become much less critical of making unmusical sounds and you can explore the musical palette of the instrument to see what lines it provokes.'' This user also observed a game-like element of the system: ``I am trying to make it say something crazy by playing crazy sounds'' and ``I am playing my guitar out of tune and try to find music in it as the machine is trying to find meaning in it.'' The user also noted that the generated lines sometimes encouraged them to stay in a certain music mode, such as minor chords, when they liked the lines that it generated. This suggests that the system could act as a responsive listener for musicians, to help them bring their music to the desired emotional expression. For example, if the musician intends their music to be sombre, and the emotions of the lines match the mood, they would know they are on track, and if the emotions in the lines start shifting as they try new expressions, it could be a signal that their music may not have the intended emotional effect.

\subsubsection{System as a jam partner.}
One user commented that the system acts like a jamming partner as they play: ``I don’t feel like I am alone. I feel like jamming with someone.''
``It's like having an uncritical jam partner. It always responds to whatever you are trying.'' ``The machine is trying to work with you even if you make a mistake.'' This suggests that a musician can benefit from using the system and be encouraged to play their instrument, even if they are not actively trying to write lyrics.

\subsubsection{System as a source of lyrics.}
Users commented that the system would be useful as a source of new lyrical ideas. User C: ``I could see it being very useful if I'm having a bit of writer's block and just noodling around to get a starting point.'' User B: ``I like to jam with my band. We jam for 30 minutes or so.  It would be nice to have it running. It would do the work for you as you are jamming. If at the end of it you have a pile of lyrics that would be great. That could be the starting material we could use for a song.'' The user emphasized that for them it would be important to see the lines after the jam session as they may miss some lines while playing their instruments: ``I look at my hands when playing, occasionally glancing at the screen to see the lyrics. I would like to see the lines after the jam session.''

\subsubsection{Generated lines and music mood.}
Users noticed differences in lines, in particular with the changes in tempo and instruments used.
One user played on their keyboard in the piano and Hammond B3 (organ) modes. The user played minor piano chords on the piano, noticing that the lines were darker in theme. Switching to  organ with a more upbeat percussive sound and using major chords more often led to more upbeat lines.

Another user noticed that the system generated interesting lines when they played unusual musical shapes: ``The system seemed to produce more interesting lines when it was fed slightly more novel audio, e.g. a tune that had a mode change (switching from major to minor)''.

The genre of music did not appear to have as strong effect as tempo or the choice of instruments. One user who played blues, rock, folk and rockabilly genres did not observe a strong effect of genre. The likely reason is that the training dataset only contains rock music, and therefore the model has never seen music of other genres. It would be interesting to see how the lines are affected by genre, if the model is trained on a larger dataset, which is more representative of different genres.  

\subsubsection{Themes in generated lyrics}
Users noted presence of recurring themes as they played. They commented on often seeing a subset of words in the lines shown to them. But these words appeared to be different for each user, which suggests that the system differentiated between their musical styles, and narrowed down on specific themes. One user commented that when they experimented by playing songs with overtly obvious emotional valence (sad and happy), the system generated lines with markedly different words: ``pain'', ``falling'', ``leaving'' for a sad song, and ``time'', ``dreams'', ``believe'' for a happy song. 

Training the model on a larger dataset would likely lead to more lexical diversity. Furthermore, the system always showed the users top two generated lines ranked by BERT, which was fine-tuned on a small dataset as well. When we started sampling from among the top-ranked lines, the diversity improved.

\section{Related Work}

Automated creative text generation has long been getting attention from researchers at the intersection of computational creativity and artificial intelligence. Creative text generation includes tasks such as story \cite{riedl2010narrative,roemmele2016writing,brahman-etal-2020-cue} and poetry \cite{manurung2004evolutionary,zhang2014chinese,yang2017generating} generation. Recently automated lyric generation has also begun receiving attention from the research community.

\citeauthor{malmi2016dopelearning}~\shortcite{malmi2016dopelearning} approached lyrics generation from an information retrieval perspective. They selected the suitable next line from a dataset containing rap lyrics to produce a meaningful and interesting storyline. Similarly, \citeauthor{yu2019deep}~\shortcite{yu2019deep} developed a technique for cross-modal learning and used it to retrieve lyrics conditioned on given audio. However, our vision with this work is not to generate complete song lyrics but individual lines that act as snippets of ideas for the artists.
%Mention lyrics-conditioned melody/song generation if there is space left.

With the recent advances in end-to-end text generation, lyrics generation is often considered a conditional text generation task.
\citeauthor{vechtomova2018generating}~\shortcite{vechtomova2018generating} generated author-stylized lyrics using audio-derived embeddings. \citeauthor{nikolov2020rapformer}~\shortcite{nikolov2020rapformer} generate rap verses conditioned on the content on any given text.
\citeauthor{savery2020shimon}~\shortcite{savery2020shimon} created a system for a real-time human and robot rap battle. They generated lyrics conditioned on the keywords present in the lyrics of the human rapper. In contrast, our model conditions the lyrics on an audio stream. %and generates lyrics that better capture the mood of the song. This is important since there are associations between chords and lyrics sentiment within songs. For example, major chords are associated with positive sentiment and minor chords with negative sentiment \cite{kolchinsky2017minor,willimek2014minor}. 

While the above systems generate lyrics conditionally, there is no focus on capturing the mood of the song while generating lyrics. This is important since there are associations between chords used in the songs and the sentiments expressed in the lyrics. For example, major chords are associated with positive sentiment and minor chords with negative sentiment \cite{kolchinsky2017minor,willimek2014minor}. 
%\citeauthor{kolchinsky2017minor}~\shortcite{kolchinsky2017minor} show that there are associations between chords and lyrics sentiment within songs. For example, major chords are associated with positive sentiment and minor chords with negative sentiment. Thus, it is important that lyrics generation systems are able to capture the mood of the song. This can be achieved by conditioning the lyrics on the song itself.
\citeauthor{watanabe2018melody}~\shortcite{watanabe2018melody} use a language model conditioned on an input melody for lyrics generation.
\citeauthor{cheatley2020co}~\shortcite{cheatley2020co} created a co-creative song writing system and applied it in a therapeutic context. The lyrics could either be fed by the user or generated by the system conditioned on an audio track or input topics selected by the user.  \citeauthor{vechtomovaLyrics2020}~\shortcite{vechtomovaLyrics2020} also developed a system to generate lyric lines conditioned on music audio. They show that their system generates lyric
lines that are consistent with the emotional effect
of a given music audio clip. In this work, we propose new approaches that align the learned latent spaces of audio and text representations, thus building upon and improving their approach. Moreover, our system generates lyrics in real-time. 

\section{Implementation Details}
% text-VAE details
\subsubsection{Text-CVAE.}
We use the Tensorflow framework~\cite{abadi2016tensorflow} to implement the text-CVAE.
The encoder is a single-layer bi-LSTM and a decoder is an LSTM. The hidden state dimensions were set to 300 and the latent space to 128. The CVAE models in our experiments were trained for 500 epochs.

% spec-VAE details
\subsubsection{Spec-VAE.}
We use PyTorch~\cite{Pytorch2019} to implement the spec-VAE.
The encoder has 4 Conv2d layers interleaved with ReLU activation function~\cite{nair2010rectified}.
The decoder is a mirror image of the encoder, with 4 ConvTranspose2d layers, interleaved with 3 ReLU activations and 1 Sigmoid activation in the end.
We use 128 latent dimensions for the mean and sigma vectors.
During training, we use a batch size of 32, learning rate of 1e-4, and Adam optimizer~\cite{kingma2014adam}.
The sampling temperature is 1.0 for both training and inference.

% GAN details
\subsubsection{GAN.} 
We use the AllenNLP library~\cite{gardner2018allennlp} to implement the GAN.
The generator and discriminator networks are 3-layered feed-forward neural networks, interleaved with ReLU activation function.
During training, we use a batch size of 32, learning rate of 1e-3, and Adam optimizer for both the generator and the discriminator.
We set $\lambda_{MSE}=1.0$ to ensure diverse lyric lines.
The sampling temperature is 1.0 during both training and inference.
Also, we train the GAN alignment network for 6 epochs.

% BERT details
\subsubsection{BERT.}
We use the Transformers library~\cite{wolf-etal-2020-transformers} to fine-tune a bert-base model for sequence classification on our custom dataset.
The model is trained for 15 epochs using a learning rate warmup scheduler for the first 500 steps of training with a weight decay of 0.01.
We use a batch size of 16 for training and 64 for inference.

\section{Conclusions}
We developed a system that generates lyrics congruent with the emotions conveyed in live instrumental music. Two novel approaches to align the learned representations of music and lyrics have been proposed: GAN-CVAE, which adversarially learns to predict the lyric representation from the music representation, and CVAE-spec, which transfers the topology of the music (spectrogram) latent space learned by the spectrogram VAE to the lyric latent space learned by the text CVAE. Our user study showed that users selected the lines generated by the proposed two methods significantly more often than the lines generated by a baseline for different types of instrumental songs. For another user study we recruited musicians performing live music. Their feedback suggested that the system could be useful not only for lyric writing, but also (a) to encourage musicians to improvise and try new musical expressions, and (b) act as a non-critical jam partner. In future work, we would like to generate lines that are not only congruent with the emotions of the music, but whose meter and phonetic structure match the beat and the composition of the song.

\section{Acknowledgments}
This  research  was  partially  supported  by  the  Natural  Sciences  and  Engineering  Research  Council  of Canada (NSERC) under grant No. RGPIN-2019-04897 and Compute Canada (www.computecanada.ca).

%\appendix{\LaTeX{} and Word Style Files}\label{stylefiles}

%The \LaTeX{} and Word style files are available on the ICCC-13
%website, {\tt http://computationalcreativity.net/iccc2013/}.
%These style files implement the formatting instructions in this
%document.

%The \LaTeX{} files are {\tt iccc.sty} and {\tt iccc.tex}, and
%the Bib\TeX{} files are {\tt iccc.bst} and {\tt iccc.bib}. The
%\LaTeX{} style file is for version 2e of \LaTeX{}, and the Bib\TeX{}
%style file is for version 0.99c of Bib\TeX{} ({\em not} version
%0.98i).

%The Microsoft Word style file consists of a single template file, {\tt
%iccc.dot}. 

%These Microsoft Word and \LaTeX{} files contain the source of the
%present document and may serve as a formatting sample.  

\bibliographystyle{iccc}
\bibliography{iccc}

\end{document}